\documentclass[reprint,aps,prx,superscriptaddress,twocolumn]{revtex4-1}
\usepackage{bm,physics,amsmath,amssymb,
	graphicx,xcolor,embedfile,comment,mathrsfs,mathtools,xfrac}
\usepackage[utf8]{inputenc}
\usepackage[normalem]{ulem}

\usepackage[T1]{fontenc}
\usepackage{tikz}

\usepackage{titlesec}
\titlespacing*{\section}{0pt}{1\baselineskip}{1\baselineskip}

\graphicspath{{figures/}}

\makeindex

\begin{document}

%\preprint{APS/123-QED}
%\title{Magneto-elastic coupling in the Kitaev candidate Na$_2$Co$_2$TeO$_6$ evidenced by anomalous spectra shift}Disentangling Coherent Phonons from Propagation Effects in the Terahertz Kerr Response of Bulk LaAlO$_{3}$
\title{\textcolor{black}{Disentangling Electronic and Ionic Nonlinear Polarization Effects in the THz Kerr Response of LaAlO$_{3}$ }
}% Force line breaks with \\
%\thanks{A footnote to the article title}%

\author{Chao Shen}
\affiliation{%
 ISTA (Institute of Science and Technology Austria), Klosterneuburg 3400, Austria
}%
\author{Maximilian Frenzel}
\affiliation{
Fritz Haber Institute of the Max Planck Society, Department of Physical Chemistry, Berlin, Germany}
\author{Sebastian F. Maehrlein}
\affiliation{
Fritz Haber Institute of the Max Planck Society, Department of Physical Chemistry, Berlin, Germany}
 \affiliation{
 Helmholtz-Zentrum Dresden-Rossendorf, Institute of Radiation Physics, Dresden, Germany
 }
  \affiliation{
 Dresden University of Technology, Institute of Applied Physics, Dresden, Germany
 }
 %\altaffiliation[Also at ]{Physics Department, XYZ University.}%Lines break automatically or can be forced with \\
\author{Zhanybek Alpichshev}%
 \email{Alpishev@ist.ac.at}
\affiliation{%
 ISTA (Institute of Science and Technology Austria), Klosterneuburg 3400, Austria
}%

%\collaboration{MUSO Collaboration}%\noaffiliation

%\author{Maximilian Frenzel}
%\author{Sebastian F. Maehrlein}
  %\email{Guru.Khalsa@unt.edu}
%
%\affiliation{
 %Depa%rtment of Physics, University of North Texas, Denton, TX 76203, USA
%}%
%\author{Delta Author}
%\affiliation{%
 %Authors' institution and/or address\\
 %This line break forced with \textbackslash\textbackslash
%}%

%\collaboration{CLEO Collaboration}%\noaffiliation

%\date{\today}% It is always \today, today,
             %  but any date may be explicitly specified

\begin{abstract}

Nonlinear responses to intense terahertz (THz) fields provide unique insights into complex dynamics of contemporary material systems. However, the interpretation of the obtained data, in particular, distinguishing genuine ionic oscillations from the instantaneous electronic responses in THz Kerr effect remains challenging. Here, we combine two-dimensional Terahertz Kerr effect (2D-TKE) spectroscopy experiments and their modeling to unravel complex THz-induced temporal oscillations in twinned LaAlO$_3$ crystals at low temperatures. We identify the 1.1 THz mode as $E_g$ Raman phonon, while 0.86 THz and 0.36 THz signals are due to spurious effects resulting from the co- and counter-propagation of THz and optical probe pulses in birefringent twin domains. Furthermore, we determine that the $E_g$ mode is excited via a two-photon process, whereas THz pulse reflections at the sample surface produce a temporal response that can mimic anharmonic phonon coupling. Our findings highlight the importance of propagation effects in nonlinear THz experiments and provide a refined framework for interpreting THz polarization dynamics in birefringent crystals.
%\begin{description}

%\item[Usage]
%Secondary publications and information retrieval purposes.
%\item[Structure]
%You may use the \texttt{description} environment to structure your abstract;
%use the optional argument of the \verb+\item+ command to give the category of each item. 
%\end{description}
\end{abstract}

\maketitle

%\keywords{Suggested keywords}%Use showkeys class option if keyword
                              %display desired

%\tableofcontents

\section{\label{sec:level1}Introduction }

The advent of table-top sources of intense terahertz (THz) pulses has made it possible to directly explore low-energy modes while keeping electronic degrees of freedom largely unaffected  \cite{liu2012terahertz,li2019terahertz,pal2021origin,song2023ultrafast,basini2024terahertz}. In high-field THz spectrocopies, intense THz pulses are used to excite low-lying degrees of freedom (e.g. phonons, magnons, etc) which can undergo nonlinear mutual interactions. The resulting real-time dynamics can be obtained by sensing the modified response of the host medium. One method developed recently is two-dimensional THz Kerr effect spectroscopy (2D-TKE), whereby two THz pulses are used to excite infrared- and Raman-active modes, which modify refractive index of the medium through Kerr effect picked up by a near-infrared probe beam \cite{johnson2019distinguishing,lin2022mapping,blank2023two}. Distinct excitation pathways of Raman-active phonons appear as characteristic patterns in the time- and frequency-domain responses, enabling direct comparison of experimental data with theoretical models \cite{johnson2019distinguishing}. However, great care must be taken when interpreting the results, making sure the initial assumptions of the model used are sound. For instance, in isotropic crystals, bulk interaction between THz and optical probe pulse can produce exponentially decaying response, that was attributed to dipole reorientation when propagation effects were ignored \cite{li2023terahertz}. In birefringent crystals, oscillatory signals that resemble coherent phonons can emerge due to phase matching effects of the THz and optical pulses \cite{frenzel2023nonlinear}. These examples illustrate the importance of propagation effects on TKE measurements that have to be taken into account when studying bulk samples in order to distinguish spurious electronic polarization effects  from the intrinsic coherent response of the medium.\\
\indent LaAlO$_3$ (LAO)  is a common wide-gap insulator ($\Delta\approx5.5$ eV) widely used as a substrate for the epitaxial growth of quantum materials. Recent single pulse TKE measurements show complex temporal oscillations at both room and low temperatures which could not be entirely accounted for using Raman spectra of LAO reported in the literature \cite{kovalev2025terahertz,basini2024terahertzparametric}. The main complication in the analysis of TKE data comes from the fact that below $T\approx813$ K LAO is characterized by rhombohedral ($D_{3d}$ point group) lattice structure  implying that it is birefringent for both THz and optical probe pulses propagating along any direction other than the optic axis. This will unavoidably give rise to rich and complex interaction dynamics between THz and near-infrared probe pulses, including non-negligible propagation effects in THz Kerr response.  \\
\indent Here, we report 2D-TKE measurements on LAO supported by detailed theoretical modeling based on four-wave mixing (FWM) simulation in an extended medium to unravel the complex third-order nonlinear THz polarization in bulk LAO samples at cryogenic temperatures. Similar to previous reports, we observe three distinct frequencies in TKE response (1.1, 0.86 and 0.36 THz). Using time-domain 2D-TKE spectroscopy we show that while $f=1.1$ THz mode shows typical features of the coherent $E_g$ Raman phonon in agreement with literature, the novel features at 0.86 and 0.36 THz only occur within specific delay ranges, indicative of propagation effect. In addition, 2D-TKE provides clear evidence that the excitation channel of $E_g$ phonon corresponds to two-photon THz absorption, while the reflection of THz pulses at the sample surface can mimic anharmonic phonon coupling. Our FWM simulations that account for the birefringence of both THz pump- and near-infrared probe pulses quantitatively reproduce all of the novel features, unambiguously distinguishing instantaneous electronic polarization from the genuine Raman response of the material.

%Additionally, LAO is inherently twinned due to thermal stresses and lattice imperfections, The phase transition is accompanied by the appearance of a Raman-active $E_g$-mode ($f\approx1$THz at room temperature) \cite{hayward2005transformation}. This mode is highly active and appears consistently in ultrafast experiments across various excitation conditions \cite{neugebauer2021comparison,hortensius2020ultrafast,kohmoto2011ultrafast}.

%Differential chopping scheme is employed to extract the nonlinear ellipticity signal:

%\begin{equation}
% \eta_\text{NL} \equiv \eta_\text{AB} - \eta_\text{A} - \eta_\text{B}    
%\end{equation}

%\noindent where $\eta_\text{A}$, $\eta_\text{B}$ and $\eta_\text{AB}$ stand for the Kerr signal induced by THz A, THz B and both THz pulses at the same time respectively \cite{lu2017coherent}.  

\begin{figure}[htbp!]
        \centering
    \includegraphics[scale=0.55]{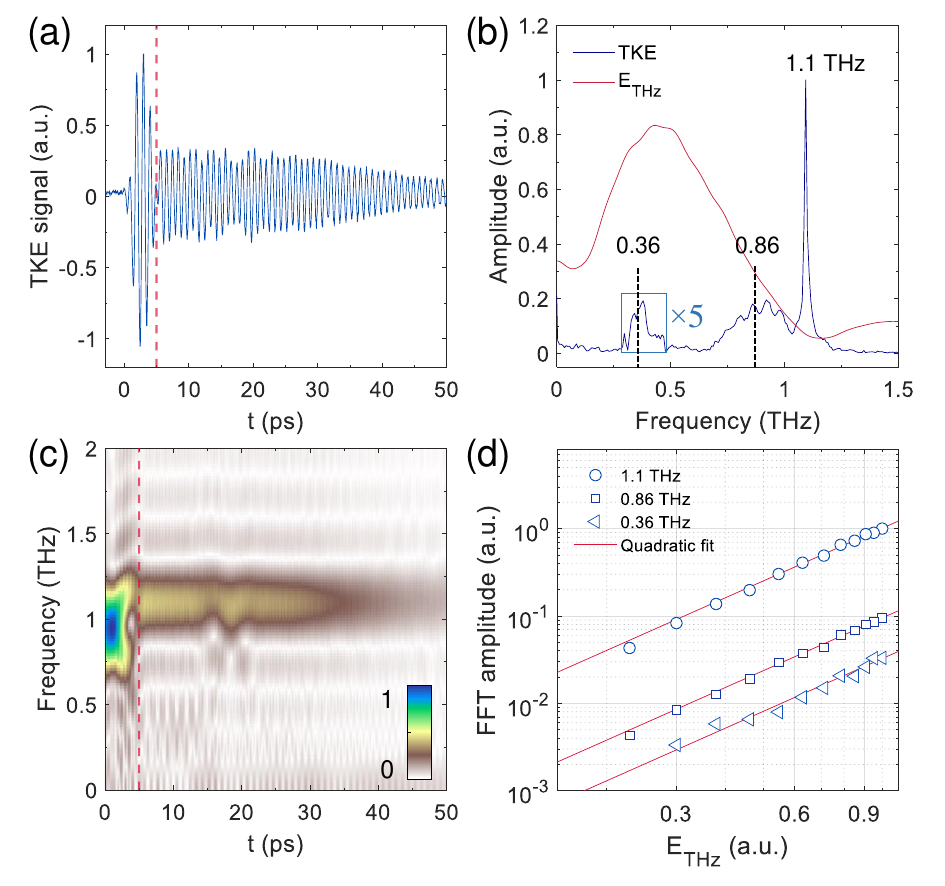}
    \caption{ Single pulse TKE measurement and field scaling of LAO at $T=5$ K. (a) Representative single pulse TKE signal at $T=5$ K. The red dashed line at 5 ps indicates the delay when THz and 800 nm probe pulses meet at the rear surface of LAO. (b) Spectra of THz pulse and Fast Fourier transform (FFT) of (a). (c) Normalized short-time FFT of (a) with a 4 ps Kaiser window. The dashed line is the same as (a). (d) Field scaling of peaks at 1.1, 0.86 and 0.36 THz.}
    \label{fig:1}
\end{figure}

\section{Results}

\subsection{Single-pulse THz Kerr spectroscopy}

In this part of the experiment we excite a [100]-cut 0.5 mm-thick LAO with an intense single-cycle THz pump pulse (peak field $E_p\approx640$ kV/cm) generated by means of optical rectification of 170 fs-long 800 nm pulses in LiNbO$_3$ \cite{hebling2002velocity, hirori2011single} and the transient Kerr effect-induced birefringence is probed by an additional 800 nm pulse arriving at a controllable delay $t$ relative to the pump. 

Fig.\ref{fig:1} (a) shows the TKE signal at $T = 5$ K. The data exhibits pronounced oscillations that start immediately at the arrival of the THz pulse. Fig.\ref{fig:1} (b) shows the Fourier amplitude of single-pulse TKE signal together with the THz excitation spectrum. Three peaks are visible here: sharp and strong mode at 1.1 THz; broad hump-like mode around $0.86$ THz and a much weaker peak at $0.36$ THz. The feature at 1.1 THz has been reported previously and was consistently identified as the $E_g$ Raman mode of LAO. We confirm this attribution by observing the intensity of the peak scaling quadratically as a function of incident THz field strength (Fig. \ref{fig:1} (d), top line). The central frequency and amplitude of the $E_g$ mode exhibit slight dependence on the lateral sample position. Since the mode frequency is associated with rotations of the AlO$_6$ octahedra about the [111] direction, this variation can be attributed to local structural variations accompanying the transition from the cubic to the rhombohedral phase \cite{salje2016direct}.

\begin{figure*}[htbp!]
        \centering
    \includegraphics[scale=0.6]{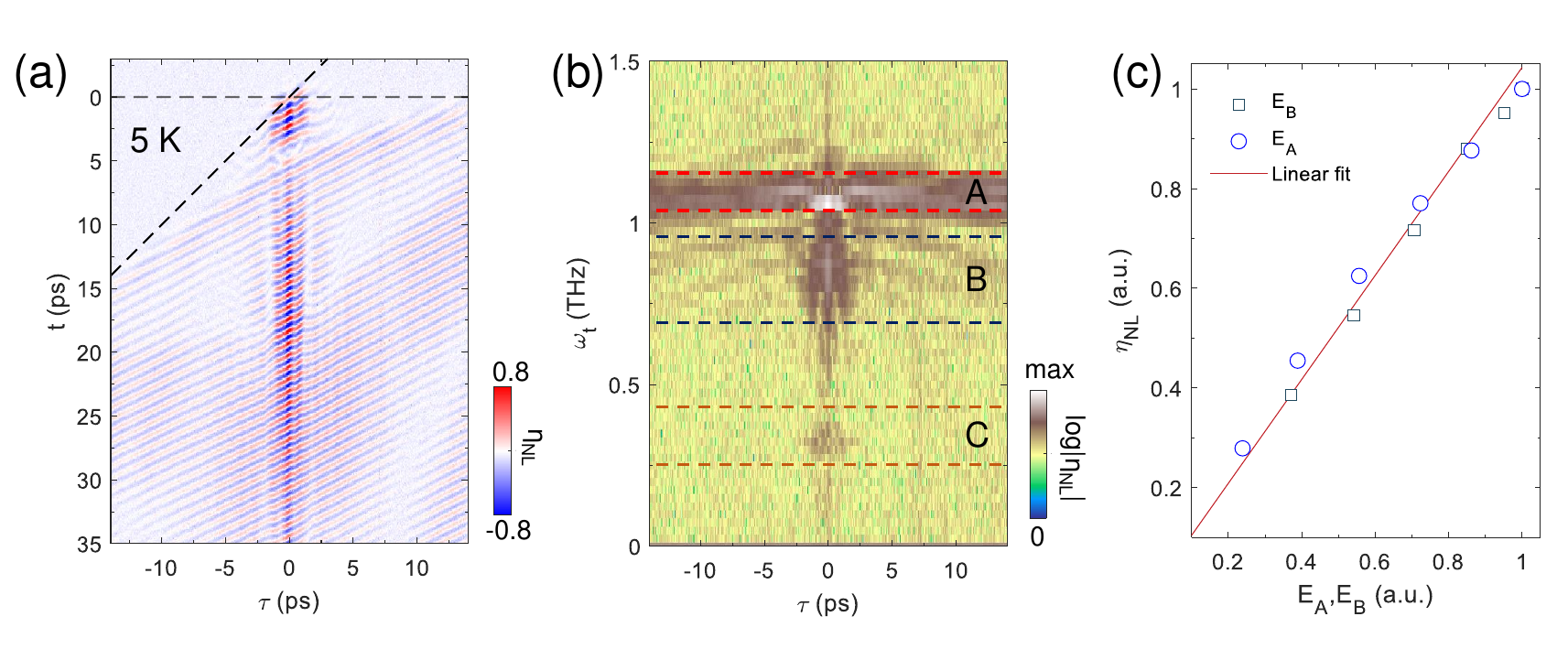}
\caption{Nonlinear 2D-TKE measurements of LAO at $T=5$ K. (a) Normalized temporal signal of nonlinear 2D-TKE. The dashed lines indicate the causality lines, namely the moment when the THz pulse arrives at the sample. (b) FFT of (a) along the $t$ axis. (c) Field scaling of nonlinear signal $\eta_\text{NL}$ at 1.1 THz as a function of $E_\text{A}$ and $E_\text{B}$, respectively. The signal is sampled along the $\tau=0$ line. In each measurement, the field of the other THz pulse was set to maximum. The fit indicates that $\eta_\text{NL}$ scales linearly on both THz pulses, ensuring a third-order response.}
    \label{fig:2}
\end{figure*}

In contrast with the 1.1 THz peak, the features at $0.86$ THz and $0.36$ THz do not correspond to any known lattice modes of the rhombohedral lattice of LAO. The first evidence that these do not reflect the intrinsic response of LAO comes already from a close inspection of the time dependence of TKE signal in Fig.\ref{fig:1} (a): we perform a short-time Fourier transform with a 4 ps-wide moving window (Fig.\ref{fig:1} (c)) and observe that unlike 1.1 THz peak, which exists for all time positions of the Fourier window, the 0.86 THz feature is only there for probe delay times less than some characteristic cut-off time $\tilde{t}\approx5$ ps. Looking back at Fig.\ref{fig:1} (a) we notice that $\tilde{t}$ is even more prominent in time domain, marking a visible change in the TKE signal. Using previously reported values for cryogenic refractive indices $n_{\text{THz}}=5$ between 0.4 and 1 THz \cite{basini2024terahertzparametric} and $n_\text{pr}\approx2$ at 800 nm \cite{Nelson2012}, we find that $\tilde{t}$ corresponds exactly to the moment when THz pump and near-infrared probe pulses meet at the rear surface of the sample, hinting that at least the 0.86 THz emerges during the spatial propagation of THz and probe pulses. This is further supported by the presence of multiple equidistant peaks on top of the 0.86 THz feature with a spacing of 0.06 THz, which can be explained by THz Fabry-Perot modes existing in a 0.5 mm thick sample \cite{Spencer2025}. In agreement with centrosymmetric symmetry of LAO which dictates that all features visible in Kerr effect result from a third-order nonlinear polarization  \cite{urban2025thz}, the intensities of both 0.36 THz and 0.86 THz features scale quadratically as a function of THz field strength (Fig. \ref{fig:1} (d)).

\begin{figure*}[htbp!]
\includegraphics[scale = 0.55]{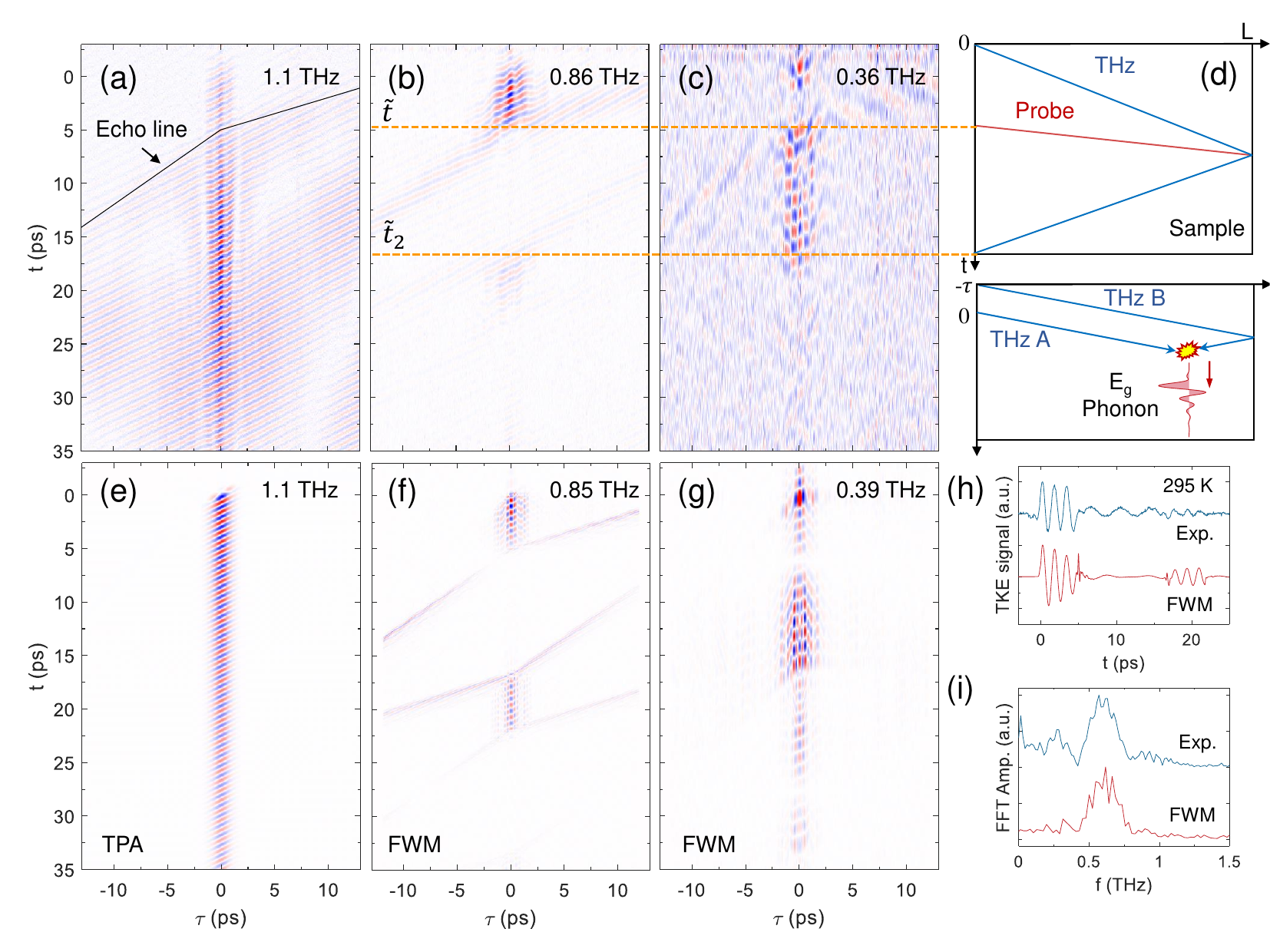}% Here is how to import EPS art
\caption{\label{fig:wide} IFFT of three peaks and modeling.  (a-c) IFFT of the 1.1, 0.86 and 0.36 THz peaks, respectively. (d) Upper panel: World line diagram at $\tau=0$, illustrating the propagation and reflection of the THz and optical probe pulses within the sample. Lower panel: World line diagram for  $\tau>0$, showing the propagation of THz pulses inside the sample. The explosion symbol marks the point where two THz pulses intersect. The collection of these points at varying $\tau$ defines the echo line shown in panel (a). (e) Simulation of 1.1 THz mode using TPA. (f-g) FWM simulation of 0.86 and 0.36 THz peaks, respectively. (h) Measured and FWM simulated single pulse TKE response at room temperature. (i) FFT of panel (h). }
\label{fig3}
\end{figure*}

\subsection{2D THz Kerr spectroscopy}

More information on the nature of the observed peaks can be revealed by two-dimensional spectroscopy. To this end, two individual THz pulses with a controllable delay $\tau$ are produced in separate LiNbO$_3$ crystals. The pump delay $\tau$ is introduced such that one of pump pulses is fixed in time (THz A) while the other one is time-delayed relative to it (THz B). A weak near-infrared pulse at 800 nm wavelength for probing comes with a probe delay $t$ relative to THz A.

Here we performed 2D-TKE measurements on LAO at $T=5$ K with both THz pump pulses polarized at 45$^\circ$ with respect to the $c$-axis of the crystal and probe polarization was aligned parallel to the $c$-axis. The measured quantity is the so-called  nonlinear ellipticity $\eta_\text{NL}$ defined as:
\begin{equation}
 \eta_\text{NL} \equiv \eta_\text{AB} - \eta_\text{A} - \eta_\text{B}    
\end{equation}
\noindent where $\eta_\text{A}$, $\eta_\text{B}$ and $\eta_\text{AB}$ stand for the Kerr signal induced by THz A alone, THz B alone, and both pulses together, respectively \cite{lu2017coherent}. The details of the experiment can be found elsewhere (see Appendix A and Ref. \cite{ShenSTO2025}).\\
\indent The nonlinear 2D-TKE signal $\eta_\text{NL}$ is shown in Fig.\ref{fig:2} (a). Here $t$ corresponds to the delay of probe pulse relative to THz A, while $\tau$ is the timing between two THz pump pulses, therefore the signal along $\tau=0$ line-cut corresponds to single-pulse TKE signal in Fig.\ref{fig:1}. 2D-TKE technique makes it possible to see how signal in Fig.\ref{fig:1} depends on the delay between two THz pulses, in particular, the spectral content of it, as shown in Fig.\ref{fig:2} (b).\\
%As we discussed above, this can be explained by the mode-beating of frequency-shifted $E_g$ phonon in a twinned region. 
\indent We first note that the 1.1 THz Raman mode is excited most efficiently when both THz pulses overlap in time ($\tau\approx0$), strongly suggesting the phonon in question is mainly excited through direct instantaneous two-photon THz absorption. We further confirm this by observing that the magnitude of the main 1.1 THz peak at $\tau=0$ scales linearly with field intensities of both THz pulses (Fig.\ref{fig:2} (c)). On the other hand, however, it is easy to notice that 1.1 THz oscillations also persist at $\tau\neq 0$ - albeit with reduced magnitude - which could be indicative of an anharmonic coupling between IR-active and Raman-active phonons that would appear as extended features in $\tau$-direction. However, a closer inspection of the data in Fig.\ref{fig:2} (a) reveals that for $\tau\neq0$ the 1.1 THz oscillations do not begin right after the arrival of the latest THz pulse, but instead start-off with a visible delay in probe time $t$, which likely implies extrinsic nature of the $\tau\neq0$ oscillations.   \\
%Interestingly, signals with large $\tau$ are not generated immediately from the causality lines (dashed lines in Fig.\ref{fig:2}(a)) but exhibit a $\tau$-dependent time delay. In the frequency domain, all modes appear in an extensive manner instead of isolated peaks, which is somehow surprising as an extended time-domain response along $\tau$ is generally the feature of anharmonic phonon coupling. We found that the spectra can be well described as line cuts along $\omega_\tau$ in the instrument response function (IRF, shown in Fig.\ref{fig:2}(c)), which is calculated by assuming instantaneous responses, namely IRF$(t,\tau) = E_\text{A}(t)E_\text{B}(t+\tau)$ \cite{mead2020sum}. As a constant response in the frequency domain indicates a Dirac function-like response in the time domain, the match between Figs. \ref{fig:2} (b) and (c) indicates that all peaks are formed instantaneously.  This conclusion is supported by the FFT along the $t$ axis (Fig.\ref{fig:2}(d)), in which both 0.86 and 0.36 THz modes only occur within the time delay equal to the THz duration (the solid blue line). The 1.1 THz mode occurs at larger $\tau$ due to THz reflections at the surface, as we will discuss below. Linear field scaling of $E_{\text{A}}$ and $E_\text{B}$ on the nonlinear signal $\eta_{\text{NL}}$ (Fig.\ref{fig:2}(e)) confirms that it is indeed the third-order response that is measured. 
\indent To clarify the mechanism behind 1.1 THz excitation as well as the physical nature of 0.36 THz and 0.86 THz peaks we trace the probe-time dependence of each of the peaks in Fig.\ref{fig:2} (b) by performing inverse Fourier transform (IFFT) of each of the segments in probe-frequency $\omega_t$ domain separately. Figs.\ref{fig3} (a), (b) and (c) show the inverse transforms of regions A, B and C in Fig.\ref{fig:2} (b) respectively.\\
\indent Fig.\ref{fig3} (a) shows that the 1.1 THz oscillations at $\tau \neq 0$ begin at what we refer to as "echo-line" (see lower panel in Fig.\ref{fig3} (d)), which for every given value of pump delay $\tau$ corresponds to probe time $t$ at which one of the THz pulses meets with the other THz pulse that was back-reflected from the rear surface of the sample \cite{ShenSTO2025}:
\begin{equation} \label{eq:5}
    t+\frac{\tau}{2}(1\pm\frac{v}{c}) = L(\frac{1}{v}-\frac{1}{c}) 
\end{equation}

\noindent where $L$ is the thickness of the sample, the $+$($-$) sign applies for negative (positive) $\tau$ and $v$ and $c$ are the group velocity of the probe and THz pulses, respectively. This suggests a very intuitive interpretation for the $\tau \neq 0$ 1.1 THz oscillations as $E_g$ Raman mode quasi-instantaneously excited by two counter-propagating THz pulses.\\
\indent Coming to the 0.86 THz and 0.36 THz features in Figs.\ref{fig3} (b) and (c), respectively, we observe that they exhibit nonphysical behavior if interpreted in terms of actual coherent phonon oscillations. Indeed, the 0.86 THz oscillation begins $t=0$ ps but ends abruptly precisely $\tilde{t}\approx5$ ps which, as mentioned above, corresponds to the probe delay at which both the pump ($\tau=0$) and probe pulse meet at the rear surface of the sample (see upper panel in Fig.\ref{fig3} (d)). Conversely, 0.36 THz oscillation (Fig.\ref{fig3} (c)) begins precisely at $\tilde{t}$ and ends as abruptly at $\tilde{t}_2 \approx 17$ ps corresponding to the probe delay when the probe pulse meets the back-reflected pump pulses at the front sample surface (see upper panel in Fig.\ref{fig3} (d)). Such abrupt changes as a function of $t$ are unmistakable signatures of propagation-caused oscillations. 
%A more detailed analysis takes into account the birefringence at both THz and near-infrared wavelengths which causes the polarizations of all pulses to change periodically during their propagation. The oscillating frequency then is determined by how many oscillations the probe experience before it overtakes the THz pulse \cite{huber2021ultrafast,maehrlein2021decoding}. The latest overlap between probe and THz occurs at the rear and front surface for the co- and counter-propagating cases and the frequencies are proportional to $n_{\text{THz}}+n_{\text{pr}}$ and $n_{\text{THz}}-n_{\text{pr}}$, respectively. The ratio between the co- and counter-propagating frequencies is then estimated to be $7/3\approx2.33$, which is consistent with experimentally measured value of $0.86/0.36\approx2.39$.

%\indent \textcolor{red}{We now provide a four-wave mixing simulation of the propagation-caused nonlinear Kerr signal.  }

\section{Discussion}
\subsection{Simulation of four-wave mixing in bulk LAO}

To achieve detailed understanding of the third-order nonlinear polarization in LAO, we utilized a FWM simulation based on the THz-THz-VIS scheme that was recently introduced to analyze the TKE in bulk samples \cite{frenzel2023nonlinear}. In this simulation, we model the third-order nonlinear polarization using the general expression:
\begin{equation}
\begin{split}
    P^{(3)}_i(t,z) = \epsilon_0 \int_{-\infty}^t dt' \int_{-\infty}^{t'} dt'' \int_{-\infty}^{t''}dt''' \tilde{R}_{ijkl} \\
    \times R(t,t',t'',t''')  E_j^{\textrm{THz}}(t',z) \\
    \times E_k^{\textrm{THz}}(t'',z)E_l^{\textrm{pr}}(t''',z)
\end{split}
\label{Eq:FWM_sim}
\end{equation}

\noindent where $z$ corresponds to the spatial axis, along the direction of propagation of the THz ($E^{\textrm{THz}}$) and probe ($E^{\textrm{pr}}$) fields. Here, $R(t,t',t'',t''')$ is the time-domain $\chi^{(3)}$ response function analog, and $\tilde{R}_{ijkl}$ captures the crystal symmetry.\\
\indent To focus on the instantaneous electronic hyperpolarizability including static birefringence and pump-probe walk-off, we model a product of temporal Dirac delta functions, $R(t,t',t'',t''')=R_{\textrm{e},0}\delta(t-t')\delta(t'-t'')\delta(t''-t''')$. For simplicity, we assume a constant refractive index in the THz pump and optical probe frequency region, where $n_{\textrm{THz}}=5$ \cite{basini2024terahertzparametric} and $n_{\textrm{pr}}=2$ \cite{Nelson2012}. In addition, we assume a constant static birefringence between the fast and slow axes of $\Delta n_{\textrm{THz}}=0.06$ and $\Delta n_{\textrm{pr}}=0.007$ in the THz and optical frequency region, which is on the order of previous estimates in LAO~\cite{kovalev2025terahertz}.\\
\indent To compute $P^{(3)}_i(t,z)$, all three contributing light fields, $E_j^{\textrm{THz}}$, $E_k^{\textrm{THz}}$ and $E_l^{\textrm{pr}}$ are propagated through the crystal on a time-space grid~\cite{huber2021ultrafast}. The computed local $P^{(3)}_i(t,z)$ acts as a source for the emitted field $E_i^{(4)}(t,z)$ via the one-dimensional inhomogeneous wave equation. $E^{(4)}(t,z)$ ultimately interferes with the co-propagating probe field $E^{\textrm{pr}}$ in a balanced detection to give rise to the Kerr effect signal. In contrast to Ref.~\cite{frenzel2023nonlinear}, we simulate a double THz pump scheme, with a variable temporal delay $\tau$ between two THz pulses.

The simulated 2D-TKE, shown in Fig.\ref{fig3} (f-g), features pronounced short-lived oscillations of 0.85~THz that end abruptly at around $t=5\textrm{ ps}$ (Fig.\ref{fig3} (f)), when slower short-lived oscillations of 0.39~THz appear, which stop at $t=16.8\textrm{ ps}$ (Fig.\ref{fig3} (g)). The simulation results agree qualitatively with the experimental measurements, and confirm the attribution of the 0.86~THz and 0.36~THz signals in the 2D-TKE to the instantaneous electronic polarization. These phase matching effects arise from the static birefringence experienced by the emitted field in co- and counter-propagation direction of the THz pump fields. Such agreements can be also found in the single pulse TKE measurement at room temperature ($\Delta n_{\textrm{pr}}=0.005$), as shown in Fig.\ref{fig3} (h-i), which further supports our assignment.

\subsection{Excitation mechanism of  $E_g$ Raman mode}

We now discuss the excitation mechanism of the $E_g$ phonon. Due to the lack of dipole moment, Raman-active phonons in centrosymmetric media can only be excited through an indirect process \cite{juraschek2018sum}, such as anharmonic coupling to IR-active phonons, two-photon absorption (TPA) \cite{maehrlein2017terahertz} and the recently proposed hybrid photon-phonon excitation \cite{khalsa2021ultrafast}. 
%In LAO, $E_g$ signal includes two contributions, the stronger localized signal and the weaker extended signal that follows the echo line. As the latter is nothing but the consequence of THz reflections, we only discuss the localized signal. 
As discussed above, the experiment points towards TPA as the most likely excitation mechanism. This is further corroborated by the fact that there is no known IR-active phonon within the bandwidth of the THz pump pulses (Fig.\ref{fig:1} (b)) \cite{abrashev1999comparative}. To elaborate on this assumption, we simulate here the temporal evolution of the $E_\text{g}$ phonon driven by a THz pulse using the following equation of motion for this mode \cite{maehrlein2017terahertz}:
\begin{equation}
      \frac{dQ_\text{R}^2}{dt^2}+2\gamma\Omega_R\frac{dQ_\text{R}}{dt}+\Omega_R^2Q_\text{R}  = \frac{\delta}{M}E(t)^2
      \label{eq:phonon}
\end{equation}

\noindent where $E(t)$ is the applied THz electric field, $M$ is the reduced mass associated with the mode, $\delta$ is the two-photon absorption coefficient, and $\gamma$ and $\Omega_R$ are the damping constant and resonance frequency of $E_g$ mode, respectively. The nonlinear Kerr signal is calculated using $\eta_\text{NL} \propto Q_{\text{R}_\text{AB}} - Q_{\text{R}_\text{A}} - Q_{\text{R}_\text{B}}$, where $Q_{\text{R}_\text{A}}$, $Q_{\text{R}_\text{B}}$ and $Q_{\text{R}_\text{AB}}$ stand for the $E_g$ phonon coordinate excited by THz A or THz B pulses alone, or with both together, respectively. We numerically solve for the differential equation Eq.\ref{eq:phonon} using the experimentally measured THz waveforms for $E(t)$. We used $\Omega_R/2\pi=1.1$ THz and $\gamma=0.005$ to match the measured Raman phonon lifetime, and  $\delta/M$ was set equal to unity for simplicity. The resulting nonlinear temporal 2D-TKE signal $\eta_\text{NL}$ calculated this way is shown in Fig.\ref{fig3}  (e) demonstrating excellent agreement with experimental 2D-TKE signal from $E_g$, save for the contribution from back-reflection of THz pulses, not included in this calculation.

\indent It is worth noted that the $E_g$ phonon was previously attributed to anharmonic coupling to a pair of THz-excited IR-active acoustic phonons and the lowest frequency mode (0.34 THz in that work) was assigned to be the result of parametric interaction between the acoustic and Raman phonons \cite{basini2024terahertzparametric}. In our model, the experimental data is naturally accounted for by the TPA and electronic hyperpolarizability without introducing additional assumptions.

\section{Conclusion}

In summary, we investigated the THz Kerr response of LaAlO$_3$ in the cryogenic regime using 2D THz Kerr spectroscopy in combination with comprehensive theoretical modeling of four-wave THz-infared mixing in bulk extended samples. Our analysis successfully reproduces the temporal dynamics and has identified the excitation mechanism of the 1.1 THz Raman $E_g$ mode as quasi-instantaneous two-photon THz absorption. In contrast, the 0.86 THz and 0.36 THz features were identified as nonlinear electronic polarizations in a thick birefringent crystal. Our findings extend beyond the specific material and provide a general framework for interpreting dynamical THz-induced response in systems where bulk propagation effects cannot be neglected.

\begin{acknowledgments}
Z.A. acknowledges support from the collaborative research project SFB Q-M\&S funded by the Austrian Science Fund (FWF, grant No.PR1050F8602). S.F.M. acknowledges support and funding from the Deutsche Forschungsgemeinschaft (DFG, grant No. 469405347). 
\end{acknowledgments}

\appendix
\section*{Appendix A. Experimental details}
\label{Appendix_Exp}
%Two individual THz pulses with a controllable mutual delay $\tau$ are produced by pumping separate LiNbO$_3$ crystals with 170fs-long ultrashort pulses at 800nm using the tilted-wavefront technique \cite{hirori2011single}. The pump delay $\tau$ is introduced such that one of pump pulses is static (THz A) while the other one is retarded relative to it (THz B). A weak near-infrared pulse at 800nm for probing coming with probe delay $t$ relative to THz A.

In the setup THz A beam passes through a 90$^\circ$ periscope in order to change its polarization from vertical to horizontal. It is then combined with (vertically polarized) THz B beam by a wire-grid polarizer. A second wire-grid polarizer is then used to clean the electric polarizations of the combined THz beam. 

The LAO crystal (MSE LLC, [100] cut) was diced into $10\times10\times0.5$ mm$^3$ samples and mounted in a liquid-helium flow cryostat with one of the sample axis parallel to the THz electric fields. A 2''-diameter 90$^\circ$-off-axis parabolic mirror with an effective focal length $F=50$ mm was used to focus the THz beam on the sample with spot radii (1/e$^2$) 552$\times$368 $\mu$m$^2$ and 586$\times$245 $\mu$m$^2$ for THz A and THz B respectively (as measured by {\it Swiss Terahertz}, RIGI micro-bolometer camera). We estimate the peak THz electric field strength $E_p$ by comparing the total energy of the pulse $W$ (measured by a pyroelectric power detector {\it GenTec} THZ5I-BL-BNC) with other beam characteristics, such as beam radius $r$, the normalized temporal shape of the pulse $g(t)$ (per electro-optic sampling with 200 $\mu$m-thick ZnTe) according to the expression \cite{ozaki2010thz}:
\begin{equation}
E_p = \sqrt{\frac{2ZW}{\pi r^2\int g(t)^2\,dt}}
\end{equation}

\noindent where $Z$ is the vacuum impedance. We estimate that peak field strengths of THz A and THz B pulses at the sample plane in the air are $E_p^A\approx440$ kV/cm and $E_p^B\approx640$ kV/cm respectively. 

Time-resolved THz Kerr effect is performed in transmission geometry with Kerr-induced ellipticity $\eta$ of the probe pulse measured by means of the standard quarter-waveplate -- Wollaston prism -- balanced photodiode combination. The incident polarization of the probe pulse was kept at $45^\circ$ relative to sample axes and the probe beam radius (1/e$^2$) on the sample was estimated to be $r_{pr}\approx200$ $\mu$m.

\section*{Appendix B. Further details on the Four-wave mixing simulation of the 2D-TKE signal}
\label{Appendix_FWM}
We use the THz-THz-VIS simulation outlined in Ref.~\cite{frenzel2023nonlinear}. The local third-order nonlinear polarization $P^{(3)}_i(t,z)$ is calculated using Eq. \ref{Eq:FWM_sim}. $P^{(3)}$ is induced by both the sum-frequency term $E_j^{\textrm{THz}}(t',z)E_k^{\textrm{THz}}(t'',z)$ (and its conjugate) and the difference frequency term $E_j^{\textrm{THz}}(t',z)E_k^{\textrm{THz},*}(t'',z)$ (and its conjugate), together with the probe field $E_l^{\textrm{pr}}(t''',z)$. All three contributing light fields are propagated through the crystal on a time-space grid as described in Refs.~\cite{frenzel2023nonlinear} and~\cite{huber2021ultrafast}. LAO is rhombohedral at room temperature and below, belonging to the $\textrm{R}\bar{3}\textrm{c}$ space group~\cite{neugebauer2021comparison}. For simplicity, we set all allowed tensor elements to have the same magnitude in $\tilde{R}_{ijkl}$: $\chi^{(3)}_{xxxx}=\chi^{(3)}_{yyyy}=\chi^{(3)}_{xxyy}=\chi^{(3)}_{yyxx}=\chi^{(3)}_{xyxy}=\chi^{(3)}_{yxyx}=\chi^{(3)}_{yxxy}=\chi^{(3)}_{xyyx}$. For the THz field $E^{\textrm{THz}}(t)$, we use the full experimental THz electric field measured using electro-optic sampling in 200 $\mu$m thick ZnTe. We assume a Fourier-limited Gaussian spectrum with a center wavelength of 800~nm and pulse duration of 20~fs for the probe field~$E^{\textrm{pr}}$. We use a time grid with a finite element size of $\Delta t'=24\textrm{ fs}$, a spatial grid with a finite element size of $\Delta z=2$ $\mu$m, and a pump-pump delay increment of $\Delta \tau=100$~fs. We consider the geometry in which the THz pump pulses are linearly polarized at $45^\circ$ with respect to the probe pulse polarization, and the probe pulse is polarized parallel to the fast axis, so that both the THz field and the emitted field experience static birefringence. Finally, the transmitted probe field $E^{\textrm{pr}}$ and the emitted field $E^{(4)}$ interfere in a balanced detection.% comprising the actions of a half-wave plate and a Wollaston prism.

\bibliography{LAO}% Produces the bibliography via BibTeX.

\end{document}